# Manifestation of Jet Quenching in the Total Transverse Energy Distributions in the Nucleus-Nucleus Collisions

M.V. Savina, S.V. Shmatov, N.V. Slavin, P.I. Zarubin

*Joint Institute for Nuclear Reseach,*

20 December, 1997

## Abstract

In the framework of the HIJING model global characteristics of nucleus-nucleus collisions are studied for a Large Hadron Collider energy scale. An interesting model prediction is the presence of a central bump over a pseudorapidity plateau of a total transverse energy distribution. The bump is induced by a jet quenching effect in a dense nuclear matter. It is shown that on a wide acceptance calorimeter with a pseudorapidity coverage $-5 < \eta < 5$ allow to obtain experimental confirmation of such an effect.

The investigation has been performed at the Laboratory of High Energies, JINR.

A dense nuclear matter formation (or quark-gluon plasma) in central heavy ion collisions at high energies is predicted by many models [1]. One of the predicted properties of such nuclear matter is parton jet energy losses as a result of final state interactions with a dense nuclear matter called jet quenching [2]. One may expect a manifestation of this effect in differential distributions of a transverse energy flow, $E_t$ in a wide pseudorapidity region. In this paper we concentrated our attention on distributions $dE_t/d\eta$ generated in the framework of the HIJING model [3]. A HIJING program steering makes it possible to switch on or to switch off a jet quenching in an easy way.

In the HIJING generator [3] soft hadron-hadron processes are simulated as classical strings with kinks and valence quark ends following the FRITIOF model [4]. Hard parton processes and the jet production are based on the well known PYTHIA program package [5]. The number of jets in inelastic pp-collisions is calculated by the eikonal approximation. The HIJING model provides a dependence of the parton structure function on a collision impact parameter.

One of the most important features of the HIJING model is the inclusion of energy losses $dE/dx$ mechanism of a parton traversing dense nuclear matter. High energy quark and gluon losses in a hot chromodynamic matter are estimated in [6]. It was shown that a dominant mechanism of energy losses is radiative ones induced by a gluon emission in soft final state interactions (bremsstrahlung). In the HIJING model this mechanism is fulfilled as consecutive transmissions of a part of the quark and gluon energy $ldE/dx$ from one string configuration to another. Energy losses per unit of path for quark and gluon jets are considered to be the same. It is supposed that the energy losses are proportional to a distance $l$ traversed by a jet after last interaction and occur only in a transverse direction within a nucleus radius $R_A$. The average free path is denoted as $\chi_s << E/\mu_d^2$, where $E$ is the hard parton energy $\mu_d$ is the Debye colour screening length. Parton interaction points are defined by a probability function:

$$dP = \frac{dl}{\chi_s} e^{-l/\chi_s}$$

An interaction proceeds until a parton jet stays inside the considered volume or the jet energy exceeds the jet production energy limit. Let's note that a space-time picture of parton shower evolution is not considered in the HIJING, and the effect of dense matter formation is introduced by a phenomenological way with parameters $\chi_s = 1$ fm and $dE/dx = 2$ GeV/fm.

The HIJING generator was used to produce 10000 events of minimal bias PbPb, NbNb, CaCa, OO interactions at a 5 TeV/nucleon collision energy. Fig. 1 a and b shows a differential distribution $dE_t/d\eta$ of the total transverse energy (i.e. hadron and electromagnetic components) over pseudorapidity $\eta$ for PbPb-collisions with and without jet quenching, respectively. A distinctive feature is the presence of the bump in central pseudorapidity ($-2<\eta<2$) region when jet quenching is switched on and absence of the bump in case when quenching is switched off. Being normalized to the transverse energy flow integral, pp distributions reproduce a AA case well enough in fragmentation regions.

We explored the collision energy dependence of an enhanced transverse energy flow. In fig.2 differential transverse energy distributions are presented for 5, 2, 1, 0.5 TeV/nucleon lead-lead collisions. One may conclude that the bump becomes distinguishable over a plateau starting just from the energy value larger than 5 TeV/nucleon. For lower energy values such an effect is not so profound due to smaller rapidity difference of fragmentation regions.

We followed a jet quenching sensitivity to a mass number for lighter colliding nuclei. Energy losses of hard parton jets depend on a parton path in a dense matter like $\Delta E(l) = l dE/dx$. Therefore reduction of the colliding nucleus radii might lead to a bump reduction in the central region with the respect to the fragmentation part. This provides an additional test of jet quenching phenomena. Fig.3 shows that lead-lead collisions demonstrate maximum quenching dependence while for CaCa and lighter ion cases the bump can be seen hardly.

Initially the observation of jet quenching was proposed as a modification of inclusive particle spectra in the central rapidity region. Fig.4 shows this effect in transverse momentum spectra. Parton energy losses in dense nuclear matter lead to an increase of the particle production at small and moderate $p_t$ and to a decrease of it at large $p_t$ simultaneously.

We note that the CMS [7] experiment with a wide calorimeter acceptance $-5 < \eta < 5$ is sufficient to observe the described modification in $dE_t/d\eta\ (\eta)$ distributions. These distributions can be obtained with a practically unlimited accuracy for various colliding nuclei by a direct summation of signals from each CMS calorimeter tower.

One of the CMS calorimeter practical problems is energy resolution depletion on boundaries of different calorimeter sections, i.e. barrel, forward, and very forward ones. We propose to smooth transverse energy spectra by normalizing nucleus-nucleus distributions to appropriate distributions for pp collisions (fig.5) with the same dependence of the energy resolution in calorimeter border regions.

It's interesting to note that jet quenching manifestation happens just in the central rapidity part leaving fragmentation regions $3<|\eta|<5$ unmodified. The latter circumstance is particularly useful for definition of nuclear collision geometry (i.e. impact parameter) independently of collision dynamics details. Besides it is useful for luminosity monitoring of various colliding ion species.

Conclusion

Using the generator HIJING we have shown that jet quenching effect can be observed in the CMS experiment calorimeter by measuring the transverse energy differential distribution in the whole CMS pseudorapidity coverage.

In a more general sense the study of the global energy flow makes it possible to define quite common rules of nucleus-nucleus collisions dynamic in ultrarelativistic energy range as well as to verify some important predictions of dense nuclear matter formation model in a sufficiently simple way.

Besides, the performed analysis allows one to conclude that the jet quenching effect is small for lighter nuclei like O, Ca enabling to investigate parton distributions in nuclei with minor destortion by final state interactions.

We would like to express our thanks to Profs. I.A. Golutvin, A.I. Malakhov, and V.N. Penev for encouraging support and stimulating discussions.

References

1. Shyryak E. – Phys. Rep. **61**, p.71, 1980.

   Gross D. et al. – Rev. Mod. Phys. **13**, p.43, 1981.

   Satz H. – Ann. Rev. Nucl. Part. Sci. **35**, p.245., 1985.

2. Gyulassy M. and Pluemer M. – Phys. Lett. **B243**, p.432, 1990.

3. Wang X. N., Gyulassy M. – Phys. Rev. **D44**, N11, p.3501, 1991;

   «HIJING 1.0: A Monte Carlo Program for Parton and Particle Production in High Energy Hadronic and Nuclear Collisions», LBL-34246, 1997.

4. Anderson B. et al. – Nucl. Phys. **B281** p.289,1987.

   Nillson-Almqvist B. and Stenlund E. – CPC 43, 387, 1987.

5. Sjostrand T. – CERN-TH 6488/92, 1992.

6. Bjorken J.D. – Fermilab Prep. 82/59 THY, 1982.

   Thoma K., Gyulassy M. – Nucl.Phys. **B351**, 491, 1991.

7. The Compact Muon Solenoid - Technical Proposal, CERN/LHCC 94-38, 15 December, 1994.

**Fig.1** The differential distribution of the total transverse energy over pseudorapidity for proton-proton collisions and lead-lead ones with (a) and without (b) jet quenching effect ($\sqrt{s} = 5\ TeV\ /\ nucl.$). The distribution are normalized on an integral transverse energy flow.

**Fig.2** The differential distribution of the total transverse energy over pseudorapidity for PbPb-collisions at energy values $\sqrt{s} = 5,\ 2,\ 1,\ 0.5\ TeV\ /\ nucl.$

**Fig.3** The differential distribution of the total transverse energy over pseudorapidity for collisions of nuclei of various mass number (Pb, Nb, Ca, O) with and without jet quenching ($\sqrt{s} = 5\ TeV\ /\ nucl.$).

**Fig.4** The proton transverse momentum distributions for central AuAu collisions at (a) RHIC ($\sqrt{s} = 200\ GeV\ /\ nucl.$) and PbPb collisions at (b) LHC ($\sqrt{s} = 5\ TeV\ /\ nucl.$) energies (-1<$\eta$<1). The solid and empty circles correspond to the distributions with and without jet quenching effect, respectively.

**Fig.5** The differential distribution of the total transverse energy over pseudorapidity for collisions of nuclei of various mass number (Pb, Nb, Ca, O) with and without jet quenching ($\sqrt{s} = 5\ TeV\ /\ nucl.$) normalized to proton-proton case.

Figure 1.

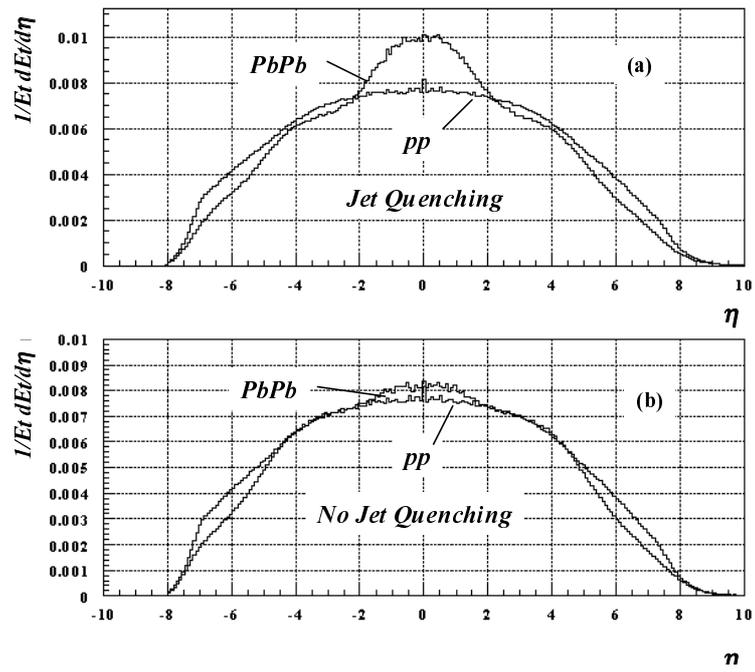

Figure 2.

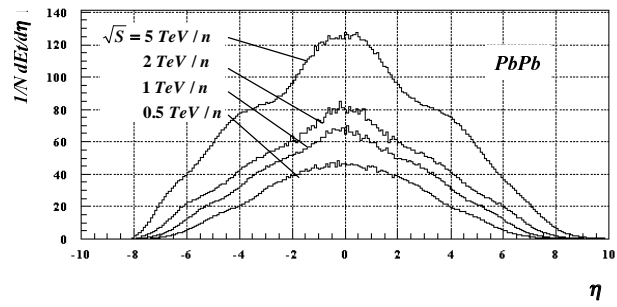

Figure 4.

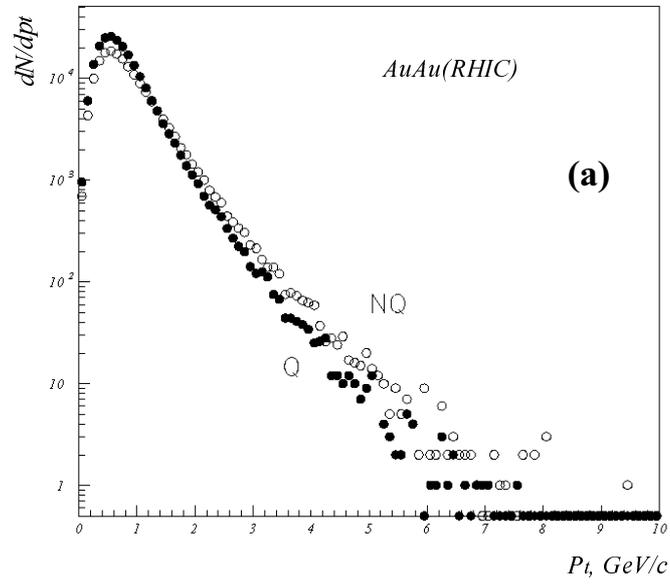

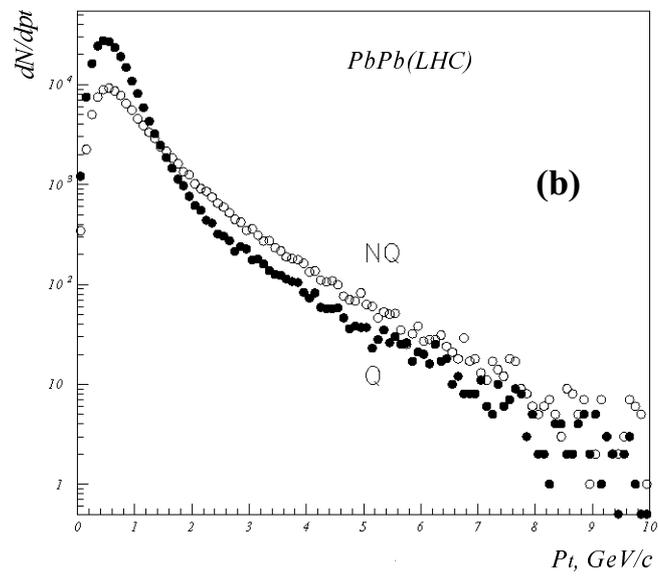

Figure 3.

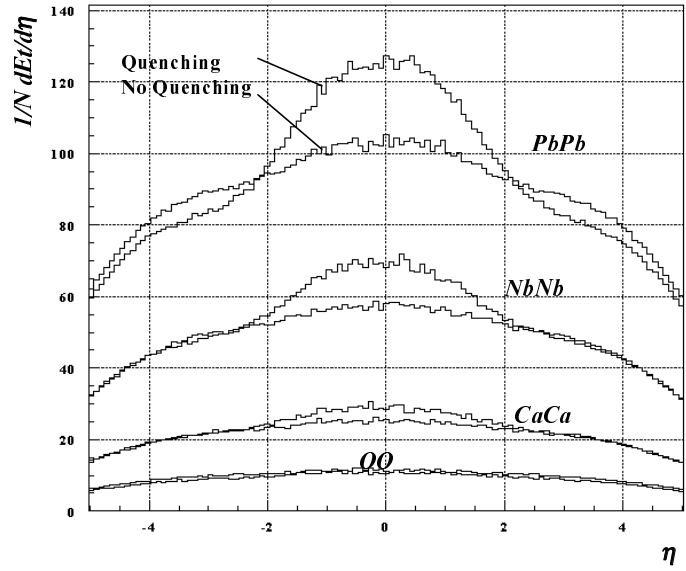

Figure 5.

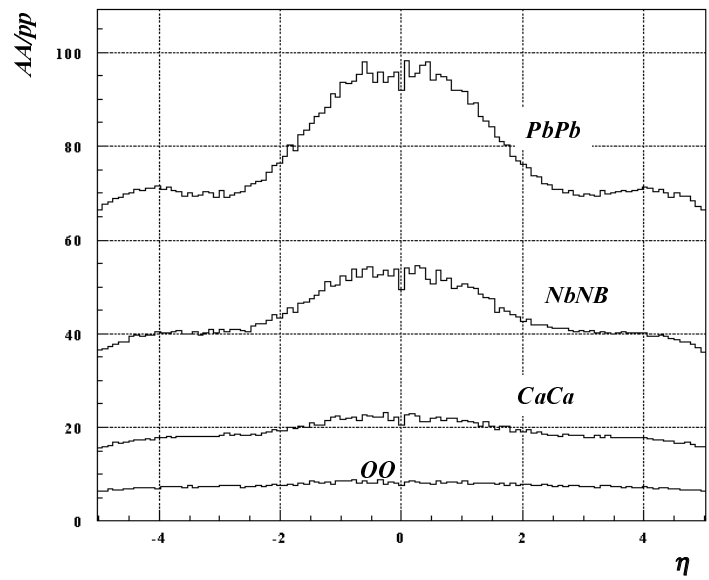